\documentclass[preprint,prb]{revtex4-1}
\usepackage{graphicx,color}
\usepackage{amssymb,amsmath,float}
\usepackage{epstopdf}
\usepackage{subfig}
\usepackage{hyperref}
\usepackage{natbib}
\usepackage{makecell}
\usepackage{multirow,diagbox,makecell,rotating}
\usepackage{multirow}
\usepackage{verbatim}
\usepackage{setspace}

\begin{document}
\author{Hong-Yue Song}
\affiliation{School of Physics and Wuhan National High Magnetic Field Center, Huazhong University of Science and Technology, 430074 Wuhan, P. R. China}
\author{Xu-Jin Ge}
\affiliation{School of Physics and Wuhan National High Magnetic Field Center, Huazhong University of Science and Technology, 430074 Wuhan, P. R. China}
\author{Man-Yu Shang}
\affiliation{School of Physics and Wuhan National High Magnetic Field Center, Huazhong University of Science and Technology, 430074 Wuhan, P. R. China}
\author{Jing-Tao L\"u}
\affiliation{School of Physics and Wuhan National High Magnetic Field Center, Huazhong University of Science and Technology, 430074 Wuhan, P. R. China}

\title{Anharmonic inter-layer bonding leads to intrinsically low thermal conductivity of bismuth oxychalcogenides}

\begin{abstract}
The anharmonicity of phonons in solid is ultimately rooted in the 
chemical bonding. However, the direct connection between phonon anharmoncity
and chemical bonding is difficult to make experimentally
or theoretically, due mainly to their complicated lattice structures. Here,
with the help of density functional theory based calculations, we discovery
that electrostatic inter-layer coupling in Bi$_2$O$_2$X (X=S,Se,Te) leads to
intrinsically low lattice thermal conductivity. We explain our discovery by the
strong anharmonic chemical bonding between Bi and chalcogen atoms. Our results
shed light on the connection between inter-layer chemical bonding and phonon
anharmonicity, which could be explored in a wide range of layered materials.  
\end{abstract}
\maketitle
\section{Introduction}
Materials with low thermal conductivity may find applications in many
disciplines including
thermoelectrics\cite{he_advances_2017,zeier_thinking_2016}, heat insulation and
phononic devices\cite{li_colloquium:_2012}. Phonons are main heat carriers in
semiconductors and insulators.  Phonon engineering of thermal conductivity has
witnessed tremendous progress in recent
years\cite{bera_marked_2010,vineis_nanostructured_2010,luckyanova_coherent_2012}.
Different extrinsic and intrinsic approaches have been developed to reduce the
phonon thermal conductivity ($\kappa$). Among them are nanostructuring, defect
engineering, and enhancing phonon anharmonicity.  Anharmonic phonon scattering
is an intrinsic mechanism that leads to finite
$\kappa$\cite{he_advances_2017,li_ultralow_2015,carrete_finding_2014,mukhopadhyay_two-channel_2018}.
Thus, utilizing strong phonon anharmonicity is an attractive way to reduce
$\kappa$, while keeping other properties intact\cite{jana_crystalline_2018}.
Generally, weak chemical bonds lead to large anharmonicity due to large atomic
displacement involved.  As a rule of thumb, complicate lattice structure with heavy elements
is believed to lead to strong anharmonicity. But their microscopic mechanism is
difficult to pinpoint.

Layered materials have weak inter-layer bonding and many show low $\kappa$ and
good thermoelectric performance\cite{zhou_promising_2017}. They normally have
good electrical transport property in the plane and low $\kappa$ across the
plane. For thermoelectric applications, it is highly desirable to reduce the
in-plane $\kappa$ ($\kappa_\parallel$) through phonon engineering.  As a kind
of thermoelectric material, layered bismuth oxychalcogenides Bi$_2$O$_2$X (X=S,
Se, Te) (BOX) have been studied
experimentally\cite{ruleova_thermoelectric_2010,luu_synthesis_2015,tan_synergistically_nodate,luu_layered_2016,zhang_synthesis_2013}.
But their thermoelectric performance is mainly hindered by the poor electrical
transport property.  Very recently, thin layer of single crystal layered
oxychalcogenide, Bi$_2$O$_2$Se, has been successfully
synthesized\cite{wu_high_2017,wu_controlled_2017}.  Its high electron mobility,
strong spin-orbit interaction and ultrafast infrared response lead to potential
applications in nano-electronics, opto-electronics, topological devices  and
ferroelectricity\cite{yin_ultrafast_2018,meng_strong_2018,wu_bismuth_2017}. The
electronic band structure of Bi$_2$O$_2$Se has been mapped out combining
angle-resolved photoemission spectroscopy and density functional theory (DFT)
calculations\cite{chen_electronic_2018}.  The experimental progress makes it
possible to enhance the thermoelectric performance by, i.e., tuning of carrier
concentration\cite{wu_low_2018}.

Despite the above mentioned progress in characterizing the electric properties,
the phonon transport properties of BOX are still poorly understood, which
hinders the in-depth understanding in its electronic, optoelectronic and thermal
properties.  Here, using  DFT based calculations, we show that,  BOX has low
intrinsic in-plane $\kappa$ ($\kappa_\parallel$). This means high electrical
conductivity and low phonon thermal conductivity can in principle be realised
simultaneously in the same direction.  Based on the analysis of its phonon
spectrum, scattering lifetime, Gr\"uneisen parameters and real-space electron
distribution, we are able to show unambiguously that, the low $\kappa_\parallel$
originates from strong anharmonic inter-layer bonding between Bi and chalcogen
atoms.  Our results shed light on the connection between chemical bonding and
phonon anharmonicity, and showed that inter-layer coupling can be used to tune
$\kappa_\parallel$ in layered materials.

\section{Results and discussions}
\subsection{Structure and phonon dispersion}
\begin{figure*}[!ht]
\includegraphics[scale=0.47]{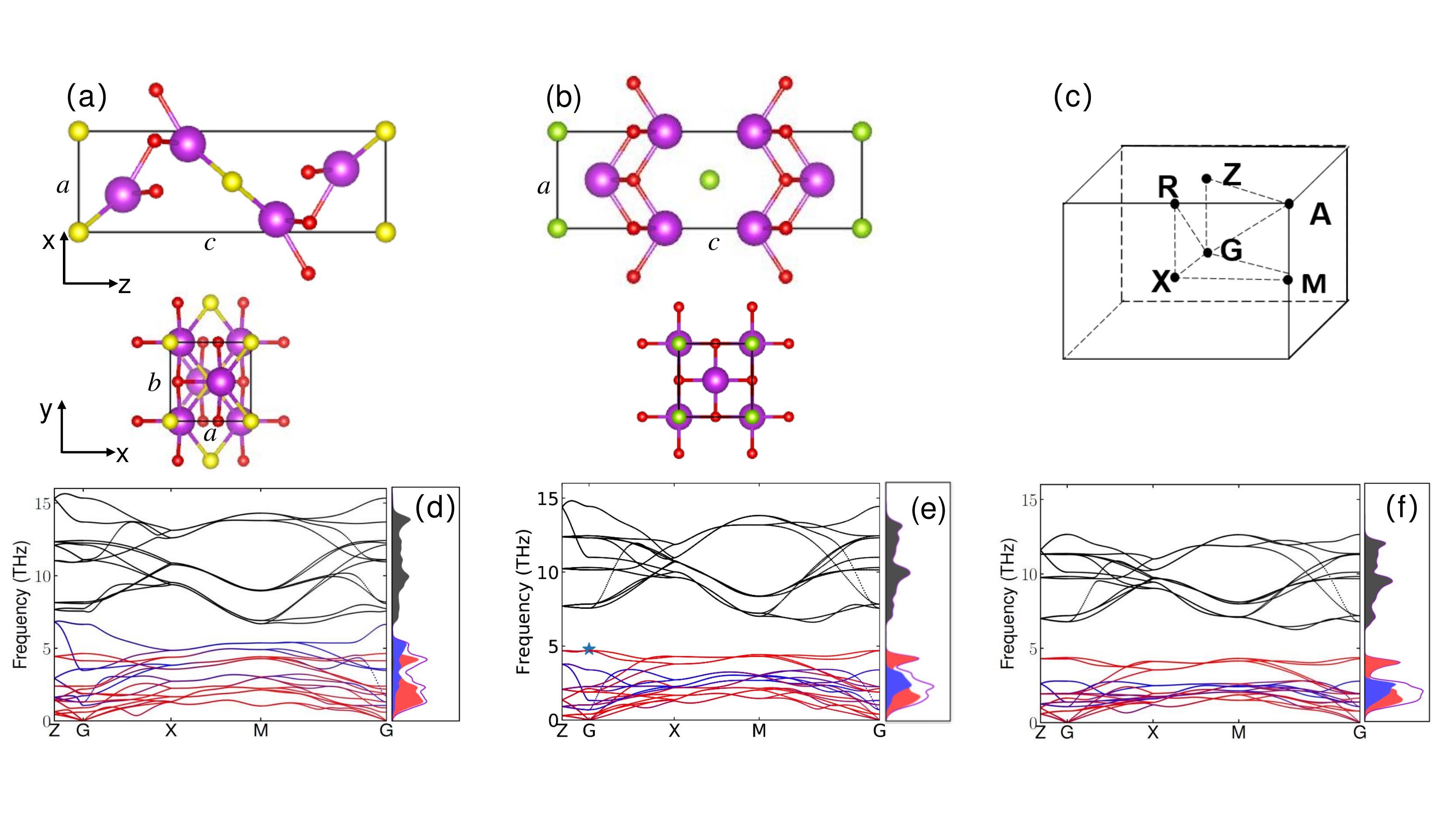}
\caption{(a-b) Side and top view of the atomic structure of Bi$_{2}$O$_{2}$X
(X=S, Se, Te). The Se and Te atom locates between two Bi$_2$O$_2$ layers, and
has 8 nearest neighbouring Bi atoms, 4 from each layer (b).  Bi$_2$O$_2$S has a
lower symmetry with slightly distorted structure, with 4 nearest neighbours, 2
from each layer (a). (c) The high symmetry points in the first Brillouin zone:
$\Gamma$ (0.0, 0.0, 0.0), X (0.5 0.0 0.0), M (0.5, 0.5, 0.0 ), A (0.5, 0.5,
0.5), R (0.5, 0.0, 0.5), Z (0.0, 0.0, 0.5). (d-f) The atomically-resolved
phonon dispersion and density of states. The red, blue and black colors
correspond to projection onto Bi, chalcogene, and O atoms, respectively. The
solid line represents the total DOS. The blue star in (e) marks the frequency
of the Raman active A$_{1g}$ mode measured
experimentally\cite{cheng_raman_2018}. The phonon bands can be divided into a
low frequency and a high frequency part. They are separated by a band gap. In
the low frequency part, Bi (red) and chalcogen (blue) atom couple together. The
high frequency part is contributed almost entirely by O atoms (black). }
\label{fig:struct}
\end{figure*}

Bulk Bi$_{2}$O$_{2}$Se and Bi$_{2}$O$_{2}$Te crystallize in the
(Na$_{0.25}$Bi$_{0.75}$)$_2$O$_2$Cl type structure, and belong to tetragonal
space group I4/mmm (139) with 10 atoms in one unit cell as shown in
Figure~\ref{fig:struct} (b).  Meanwhile, Bi$_2$O$_2$S has a distorted
structure, where the Bi atoms slide slightly apart, and belongs to the Pnnm
(58) group with lower symmetry (Figure~\ref{fig:struct} (a)). This structure
distortion changes the chemical bonding environment of S and brings a small
anisotropy between $x$ and $y$ direction. We will show below that it has
important effect on $\kappa$. There are two electron transfer from the
Bi$_2$O$_2$ layer to chalcogen layer, and the two layers are bonded through
electrostatic force.  The optimized lattice parameters, the dielectric constant
and Born effective charge of each atom are listed in Table 1 and 2 of the
Supporting Information (SI). We have also calculated the electronic band
structures from the optimized lattice parameters. They show good agreement with
previous works (SI, Figure 1).

Figure~\ref{fig:struct} (d-f) show the atomically resolved phonon dispersion
and the corresponding density of states (DOS) projected onto different atoms.
We observe a phonon gap at $\sim 6$ THz, separating the spectrum, albeit small
in Bi$_2$O$_2$S. The high frequency part of the spectrum is contributed
dominantly by O atoms. The corresponding dispersion and DOS are very  similar
for all three materials.  Although O and Bi atoms are strongly bonded in the same layer, their
motions are decoupled due to large mass mis-match. On the other hand, although
Bi and chalcogen atoms are at different layers, they couple together and form
the low frequency phonon band. This clear separation of O vibration from others
makes our analysis easier.

\subsection{Thermal conductivity}
\begin{figure}[!ht]
\includegraphics[scale=1.5]{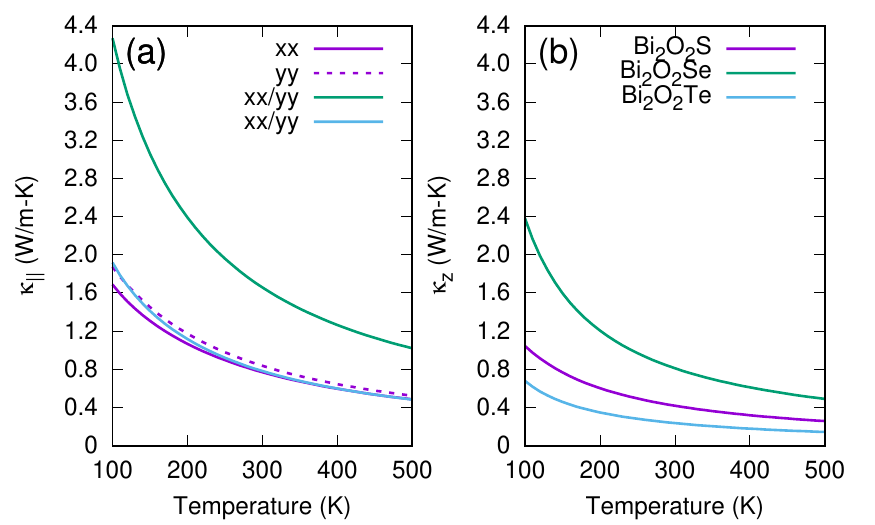}
\caption{The in-plane (a) and out-of-plane (b) phonon thermal conductivity as a
function of temperature. The layered structure in Bi$_2$O$_2$X leads to
anisotropy in $\kappa$, i.e., $\kappa_\parallel > \kappa_z$.}
\label{fig:kappa}
\end{figure}

We calculate  $\kappa_{}$ using the Boltzmann transport equation (BTE) within
the single mode relaxation time approximation (RTA) as implemented in
Phonopy\cite{phonopy} and Phono3py\cite{phono3py} package. The results are
shown in Figure~\ref{fig:kappa}. More details of the results are shown in
Figures 2-4 of SI. Consistent with their layered structure, all materials show
anisotropic $\kappa$, with $\kappa_\parallel > \kappa_z$. Bi$_2$O$_2$S and
Bi$_2$O$_2$Te show similar low  $\kappa_{x/y}$, which is below 1 W/m-K at T=300
K, while Bi$_2$O$_2$Se has around two-fold larger value. We note that,
$\kappa$ we obtained for BOX is comparable to values of other thermoelectric
materials, including
SnSe\cite{carrete_low_2014,xiao_origin_2016,li_orbitally_2015}, Bi$_2$Te$_3$,
PbX (X=S, Se, Te) and BiCuOX (X=S, Se,
Te)\cite{lee_resonant_2014,ji_low_2016,shao_first-principles_2016}. Due to
their similar lattice structure, we have performed detailed comparison to
BiCuOX in the SI (Figures 5-9 and Table 3). Additionally, we notice that
Bi$_2$O$_2$S shows lower $\kappa$ than Bi$_2$O$_2$Se. Normally, from S to Te,
we expect $\kappa$ to  decrease with increasing atomic mass for similar lattice
structures. We may attribute this abnormal behavior of Bi$_2$O$_2$S to its
structure distortion. The purpose of the rest analysis is to give a microscopic
explanation of the above mentioned features in $\kappa$.

As shown in Eq.~(\ref{eq:kappaw}) of Methods, within all the parameters that
$\kappa$ depends, the relaxation time $\tau_\lambda$ is determined by
anharmonic phonon scattering, while all the rest terms are determined by the
harmonic phonon spectrum. To find the origin of low $\kappa$ in Bi$_2$O$_2$X,
firstly we set $\tau_\lambda =1$, and show the accumulative sum of frequency
dependent $\kappa(\omega)$ (Eq.~\ref{eq:kappawac} of Methods) in
Figure~\ref{fig:jdos} (a). Bi$_2$O$_2$S shows the largest value, opposite to
the results of $\kappa$.  This means lower $\kappa$ of Bi$_2$O$_2$S does not
come from the harmonic phonon spectrum, but from their shorter $\tau_\lambda$.
We have also shown a comparison to BiCuOX (X=S, Se, Te) in Figure 7 of the SI.
We find that, although they have similar $\kappa$, the band structure
contribution to $\kappa$ in BOX is much larger than BiCuOX. This suggests
stronger anharmonic scattering and hence smaller $\tau_\lambda$ in BOX, which
we focus on in the following.

\subsection{Anharmonic scattering}
Figure~\ref{fig:knotau} shows the distribution of scattering lifetime $\tau$ as
a function of frequency for the three materials considered. The high frequency
part shows little frequency dependence  and is of similar magnitude for all
materials. The low frequency part has a much wider distribution. But it is
still clear that, the data of Bi$_2$O$_2$Se locate in higher $\tau$ region
compared to Bi$_2$O$_2$Te and Bi$_2$O$_2$S. This confirms the expectation that,
lower $\kappa$ of Bi$_2$O$_2$S and Bi$_2$O$_2$Te originates from its shorter $\tau$ and hence
stronger anharmonic scattering. Considering their similar lattice structure,
shorter $\tau$ of Bi$_2$O$_2$Te compared to Bi$_2$O$_2$Se can be understood due
to larger mass of Te. However, this can not explain why Bi$_2$O$_2$S shows the
lowest $\tau$ distribution in the low frequency regime. Thus, we need further
analysis of $\tau$.

\begin{figure}[!ht]
\includegraphics[scale=0.55]{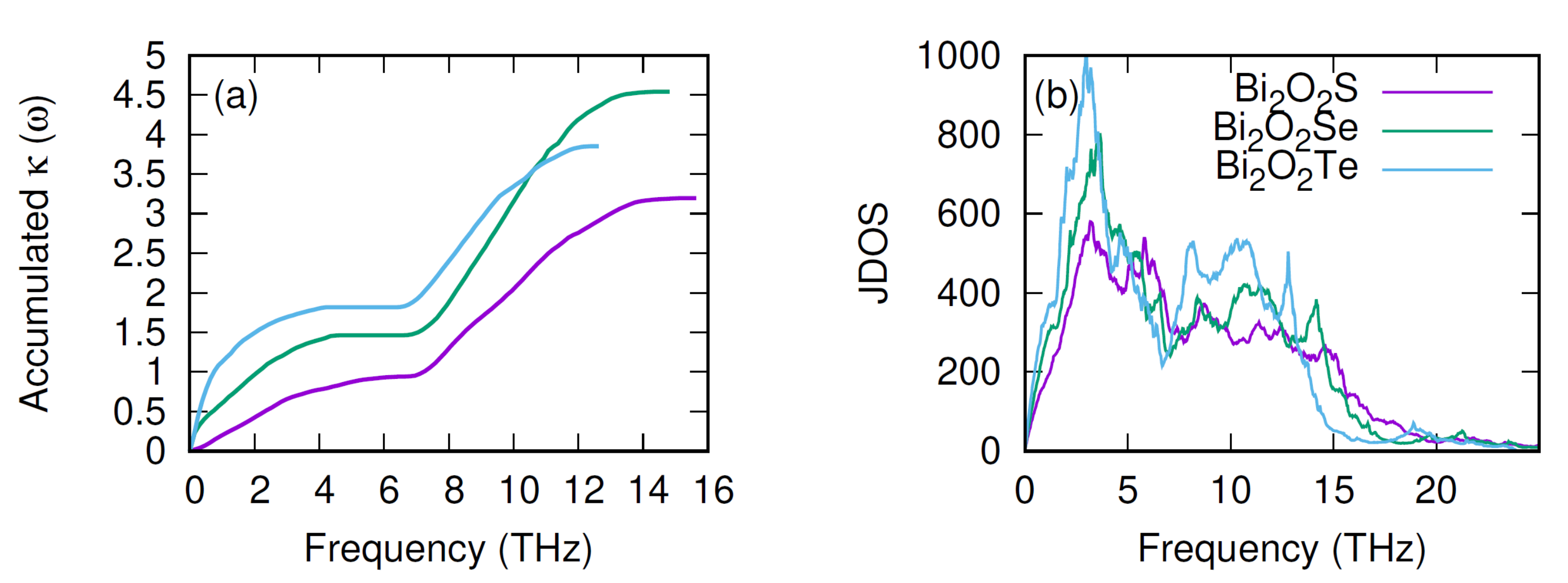}
\caption{(a) The relative magnitude of phonon band structure contribution to
the accumulated phonon thermal conductivity. (b) The joint density of states
(JDOS) for 3-phonon scattering as a function of frequency.  }
\label{fig:jdos}
\end{figure}

\begin{figure*}[!ht]
\includegraphics[scale=0.34]{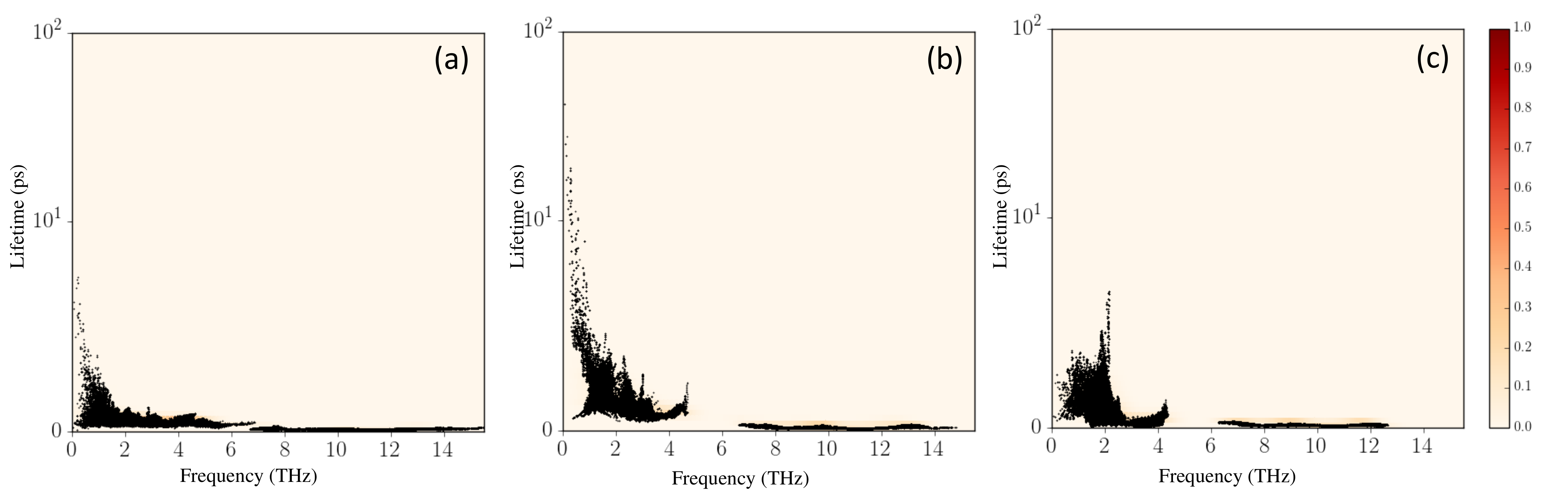}
\label{fig:tau}
\caption{Distribution of relaxation time as a function of frequency for (a)
Bi$_2$O$_2$S, (b) Bi$_2$O$_2$Se, (c) Bi$_2$O$_2$Te, respectively.
}
\label{fig:knotau}
\end{figure*}

The magnitude of $\tau$ depends both on the scattering phase space and on the
strength of anharmonic potential. We can separate the two contributions.
Figure~\ref{fig:jdos}(b) compares the joint DOS (JDOS) of 3-phonon scattering
in all materials. We find that with increasing atomic number from S, Se to Te,
JDOS in the low frequency regime grows up. The reason is that, heavier mass
leads to narrower phonon spectrum and stronger overlap between different modes.
This tends to increase the phase space for anharmonic phonon scattering.
Although the change of JDOS may partly explain different $\kappa$ of
Bi$_2$O$_2$Se and Bi$_2$O$_2$Te, it still can not explain the lower $\kappa$ of
Bi$_2$O$_2$S. 

We now turn to the anharmonic potential, the strength of which can be
characterized by the Gr\"uneisen parameters ($\gamma$). To further correlate
with the atomic chemical bonding, in Table~\ref{tab:gru}, we show the projected
$\gamma$ on different atoms in three directions. As a common feature, we get
larger $\gamma$ for Bi and chalcogen atoms compared to O atoms. These results
suggest that inter-layer Bi-X bonding in BOX is strongly anharmonic and
generates stronger phonon scattering. This is the common feature of all three
kinds of materials.  The low $\kappa$ of Bi$_2$O$_2$S can also be understood
from the Gr\"uneisen parameter. The lattice distortion in Bi$_2$O$_2$S makes
the $x$ and $y$ direction anisotropic. This is reflected in the Gr\"uneisen
parameters. $\gamma$ of S in $y$ direction is reduced, while that in $x$ and
$z$ gets much larger, promoting the inter-layer anharmonicity. This information
suggests that, we can attribute the smaller scattering lifetime and lower
$\kappa$ of Bi$_2$O$_2$S to the enhancement of anharmonic inter-layer coupling
generated by lattice distortion.

Further comparison to BiCuOX supports our above arguments. In BiCuOX, chalcogen
atom and Cu form stronger bonds. This reduces the anharmonic inter-layer
coupling between BiO and CuX layers. This is reflected in the projected
Gr\"uneisen parameters. First, in BiCuOX, $\gamma$ in $z$ direction is
consistently smaller than that of in-plane ($x/y$) direction for all atoms. Second, Cu
atoms show the largest $\gamma$ instead of chalcogen atoms in BOX. The
reduction of $\gamma$, together with their smaller JDOS (SI,
Figure~{8}), leads to much longer scattering lifetime (SI,
Figure~{9}).

\begin{figure}[!ht]
\includegraphics[scale=0.45]{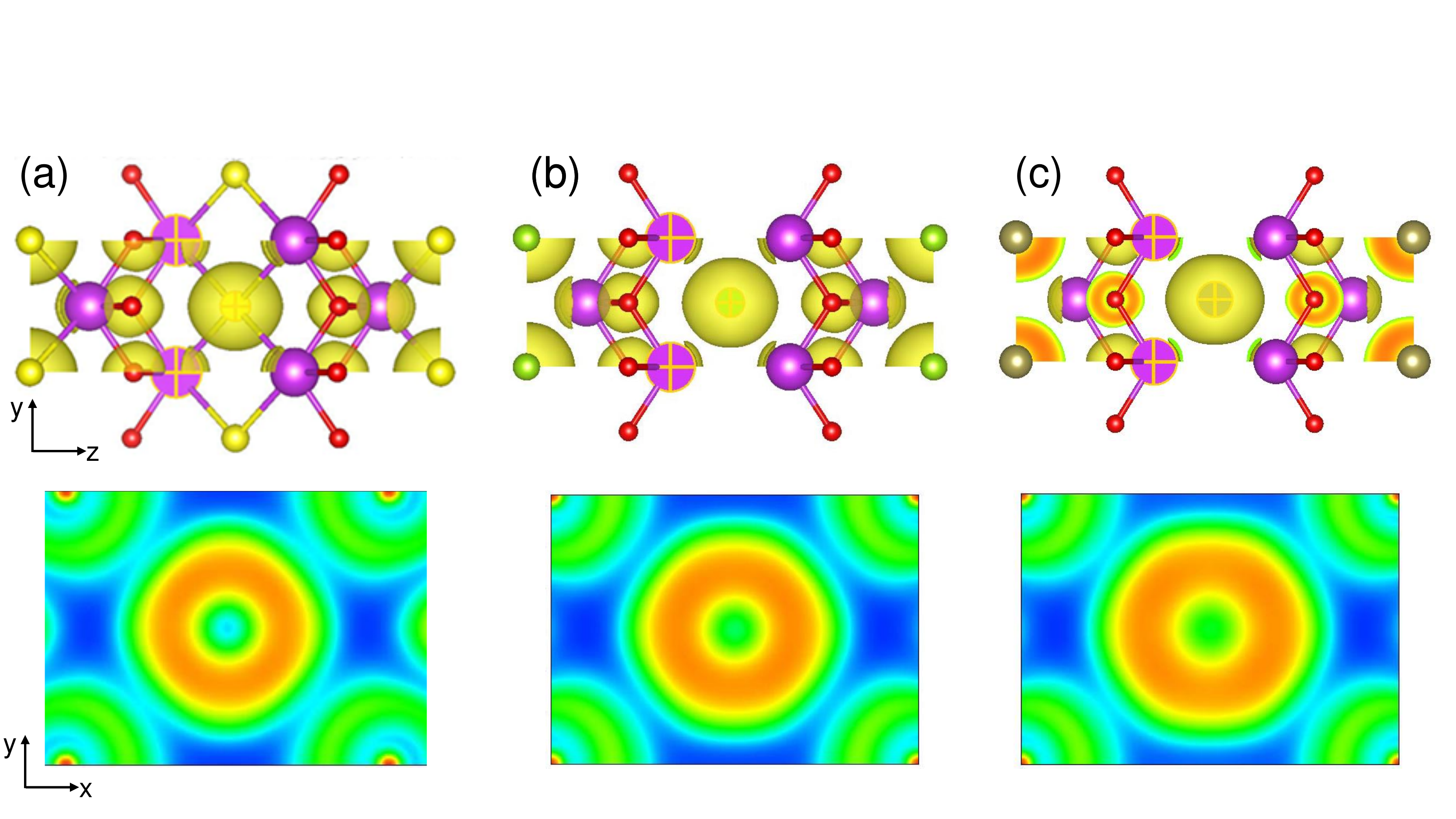}
\caption{Electron localization function (ELF) of (a) Bi$_2$O$_2$S, (b)
Bi$_2$O$_2$Se  and (c) Bi$_2$O$_2$Te. The upper panel shows the atomic
structure and the isosurfaces at 0.6. The LPE of Bi is seen as caps near Bi
atoms. The lower panel shows the plane cut defined by the three atoms marked by
crosses. The ELF is plotted from 0 (blue) to 1 (red). Comparing the
distribution in the lower panels, we can see that the ELF distribution spreads
further out, which means that the lattice distortion enhances the inter-layer
coupling between S and Bi.}
\label{fig:elfcut}
\end{figure}

\begin{table*}[htbp]
\begin{tabular}{cc|ccccccccccccc}
\hline
\hline
 \diagbox{material}{$\gamma$}& &Bi(x) & Bi(y)&Bi(z) &  X(x)&X(y) & X(z) &  O(x)&O(y) &O(z)\\
\hline
Bi$_{2}$O$_{2}$S& & 4.79 & 4.79 & 4.02 & 4.70 & 2.40 & 4.93 & 1.30 & 1.43 & 1.48\\
Bi$_{2}$O$_{2}$Se& & 4.27 & 4.27 & 3.42 & 4.05 & 4.05 & 4.34 & 1.96 & 1.96 & 1.93\\
Bi$_{2}$O$_{2}$Te& & 3.60 & 3.60 & 3.07 & 3.76 & 3.76 & 5.20 & 1.71 & 1.71 & 1.50\\
\hline
\hline
\end{tabular}
\caption{Projected Gr\"uneisen parameters $\gamma$ on different atoms and directions in Bi$_{2}$O$_{2}$X.}
\label{tab:gru}
\end{table*}

\subsection{Origin of the strong anharmonicity}
From the above analysis, we can attribute low $\kappa$ of BOX to large
anharmonic coupling between Bi and chalcogen atoms. The lattice distortion of
Bi$_2$O$_2$S further enhances the anharmonicity. The next question we ask is
what is the microscopic mechanism of the strong anharmonicity. We will now make
connection to its atomic structure and chemical bonding. In BOX, each chalcogen
atom is surrounded by a Bi cage with eight Bi atoms. Due to the inter-layer
charge transfer, the chalcogen atom interacts mainly through electrostatic
force with surrounding Bi atoms, forming weak ionic bonds. Their bonding
environment can be characterized through the electron localization function
(ELF)\cite{beck_simple_1990} (Figure~\ref{fig:elfcut}), which shows clear ionic
bonding between atoms. The small caps near Bi represents its lone-pair
electrons (LPE), which also contribute to the inter-layer coupling, i.e., they
interact through electrostatic force with the chalcogen atoms. As for
Bi$_2$O$_2$S, after lattice distortion, four Bi atoms move even closer S. This
further enhances their mutual interaction, which  can be seen from the plane
cut in the lower panel of Figure~\ref{fig:elfcut} (a). Comparing to (b) and
(c), the strongest bonding between S and Bi can be deduced. This is the
microscopic reason why Bi$_2$O$_2$S has $\kappa$ as low as Bi$_2$O$_2$Te and
much lower than Bi$_2$O$_2$Se. 

\section{Conclusions}
To conclude, we have predicted low phonon thermal conductivity of bismuth
oxychalcogenides Bi$_2$O$_2$X (X=S, Se, Te). Through careful analysis of their
phonon properties, we can ultimately correlate the strong anharmonicity and
hence low thermal conductivity to the inter-layer bonding between Bi and
chalcogen atoms. The strong correlation between bond anharmonicity and low
thermal conductivity gives atomic insights of the thermal properties of
materials. The same principle can be applied to a broad range of layered
materials with electrostatic inter-layer coupling.

\section{Methods}
\label{sec:method}
\subsection{DFT calculations}
For the  first-principles calculations, we use density functional theory (DFT)
with the projected augmented wave (PAW) method as implemented within the Vienna
$ab$ $initio$ Simulation Package
(VASP)\cite{kresse_efficient_1996,kresse_efficiency_1996}. We choose  the
Perdew-Bueke-Ernzerhof (PBE)\cite{perdew_generalized_1997} version of
generalized gradient approximation (GGA) to treat the exchange-correlation
interaction. The plane wave  cut-off energy is set as $550$ eV. The Brillouin
zone is sampled by using the Monkhorst-Pack scheme\cite{monkhorst_special_1976}
with $9\times9\times3$ mesh  k-points to optimize the  structure until the
forces on the atoms are less than $0.01$ eV/\AA. 

To calculate the phonon dispersion and phonon conductivity, we use
phonopy\cite{phonopy} and phono3py\cite{phono3py} codes together with
VASP\cite{kresse_efficient_1996,kresse_efficiency_1996}. The second and third
order force constants are calculated  by finite-difference method. We use
$5\times5\times2$ supercell for the second order force constant, and  $3\times
3\times2$ supercell for the third order force constant. We have also performed
the $4\times4\times 2$ supercell calculation to confirm the convergence of the
third order forces. We set the convergence
criteria to $10^{-8}$ eV for self-consistent loop and  $0.01$ eV/\AA~for the
force. The $\Gamma$-only scheme is used to sample the reciprocal cell of the
supercell. 

\subsection{Thermal conductivity}
The linearized phonon BTE with the single mode RTA is used to calculate the
phonon thermal conductivity with $21\times21\times21$ sampling mesh using
phono3py\cite{phono3py}. We first define
\begin{equation}
\vec{\kappa}_\lambda = C_\lambda \vec{v}_\lambda \otimes \vec{v}_\lambda \tau_{\lambda},
\end{equation}
as contribution of each mode $\lambda$ to the thermal conductivity. Here, $C_\lambda$ is the heat capacity of mode $\lambda$, $\vec{v}_\lambda$ is the group velocity, and $\tau_\lambda$ is the relaxation time. The frequency-resolve version is then written as
\begin{equation}
\vec{\kappa}(\omega) = \sum_\lambda \delta(\omega-\omega_\lambda) \vec{\kappa}_\lambda.
\label{eq:kappaw}
\end{equation}
The accumulated sum plotted in Fig.~\ref{fig:knotau} is calculated as 
\begin{equation}
\vec{\kappa}_s(\omega)=\sum_{\omega_\lambda \le \omega} \vec{\kappa}_\lambda,
\label{eq:kappawac}
\end{equation}
with $\tau_\lambda$ set to 1.
Finally, the thermal conductivity $\kappa$ is expressed as
\begin{equation}
    \vec{\kappa} = \frac{1}{V} \sum_\lambda \kappa_\lambda,
    \label{eq:kappa2}
\end{equation}
where $V$ is the volume of the system.  \subsection{The joint density of
states} The joint density states can be used to quantify the phase space for
phonon anharmonic scattering, determined by the phonon dispersion relation. The
results shown in Fig.~\ref{fig:knotau} (b) is for 3-phonon scattering and
calculated from
\begin{eqnarray}
    D(\omega) &=& \frac{1}{N}\sum_{\lambda_1,\lambda_2}[\delta(\omega+\omega_{\lambda_1}-\omega_{\lambda_2})+\delta(\omega-\omega_{\lambda_1}+\omega_{\lambda_2})] \nonumber\\
    &+&\frac{1}{N}\sum_{\lambda_1,\lambda_2}\delta(\omega-\omega_{\lambda_1}-\omega_{\lambda_2}).
\end{eqnarray}

\subsection{The projected Gr\"uneisen parameter}
The mode-resolved Gr\"uneisen parameters are calculated within the
quasi-harmonic approximation using Phonopy.  To characterize the anharmonicity
of each atom, we define the projected Gr\"uneisen parameter by projecting all
the modes to given atom in a given direction as following
\begin{equation}
\gamma_{i\alpha} =  \frac{\sum_\lambda \gamma_\lambda e_{\lambda,i\alpha} }{\sum_\lambda e_{\lambda,i\alpha}}.
\end{equation}
Here, $\gamma_\lambda$ is the Gr\"uneisen parameter of mode $\lambda$, $i$ is
the atom index, $\alpha$ represents the direction, and $e_{\lambda,i\alpha}$ is
the element of the eigen vector $e_{\lambda}$ corresponding to atom $i$ in
direction  $\alpha$.

%

\begin{acknowledgements}
The authors are supported by the National Key Research and Development Program
of China (Grant No. 2017YFA0403501), the National Natural Science Foundation of
China (Grant No. 21873033) and the program for HUST academic frontier youth
team. They thank the National Supercomputing Center in Shanghai for providing
computational resources. 
\end{acknowledgements}

\appendix
\section{Electronic structure}
\begin{figure*}[ht!]
\includegraphics[scale=0.7]{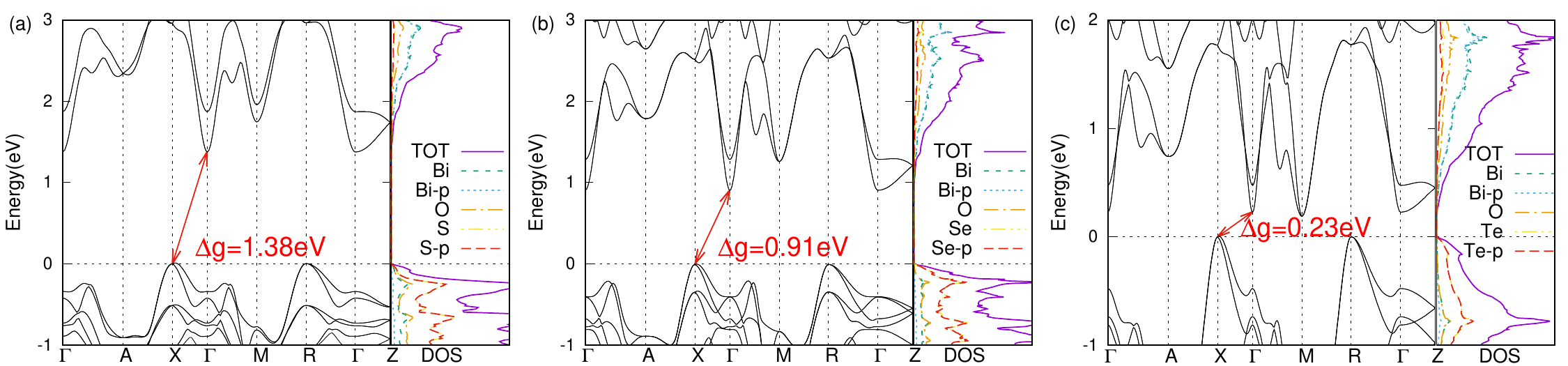}
\caption{Band structure, atomic and orbital decomposition of density of states of bulk Bi$_{2}$O$_{2}$S (a), Bi$_{2}$O$_{2}$Se (b),  Bi$_{2}$O$_{2}$Te (c). }
\label{fig:eband}
\end{figure*}

Based on the relaxed lattice structure, the electronic band structure is calculated in the modified Becke-Johnson meta-GGA potential\cite{becke_simple_2006,tran_accurate_2009} including the spin-orbit interaction. The results are shown in Figure~\ref{fig:eband}. 
We get a band gap of 1.38 eV, 0.91 eV and 0.23 eV for Bi$_{2}$O$_{2}$S, Bi$_2$O$_2$Se and Bi$_2$O$_2$Te, respectively. They are comparable to experimental observations.
The DOS plots show that at the top of the valence band, the contribution of S/Se/Te atom is dominant, while at the bottom of the conduction band the contribution of Bi atoms is dominant. Further analysis shows that, for all cases, the contribution mainly comes from the $p$ orbitals.

\begin{table}[h!]
\begin{tabular}{lcccccccccccccc}
\hline
& $a$ (\AA)& $b$ (\AA)& $c$ (\AA) & $\varepsilon$$_{x/y}$ &$\varepsilon$$_{z}$ \\
\hline
Bi$_{2}$O$_{2}$S & 3.98&3.89&12.08&10.02/11.36&9.42\\
Bi$_{2}$O$_{2}$Se & 3.93 & 3.93 & 12.40 &13.86/13.86 & 10.34 \\
Bi$_{2}$O$_{2}$Te&4.02&4.02 & 12.88 & 17.21/17.21& 12.00 \\
\hline
\end{tabular}
\caption{Calculated lattice parameters, static dielectric tensor ($\varepsilon$) of BOX.}
\label{tab:lattice}
\end{table}

\begin{table}[h!]
\begin{tabular}{lccccccccc}
\hline
& $z_{\rm Bi,x/y}$ & $z_{\rm Bi,z}$  & $z_{\rm X,x/y}$ & $z_{\rm X,z}$ &  $z_{\rm O,x/y}$ &$z_{\rm O,z}$\\
\hline
Bi$_{2}$O$_{2}$S & 5.17/5.91&5.17&-3.30/-4.32&-3.26&-3.50/-3.75&-3.54\\
Bi$_{2}$O$_{2}$Se & 6.12/ 6.12&5.39 &-4.33/-4.33&-3.12&-3.96/-3.96&-3.82 \\
Bi$_{2}$O$_{2}$Te& 6.54/6.54& 5.67 & -4.35 /-4.35 & -3.21 & -4.33/-4.33 & -4.05 \\
\hline
\end{tabular}
\caption{Calculated Born effective charge ($z$) for ions in the unit of $e$ for BOX.}
\label{tab:born}
\end{table}

\begin{table*}[htbp]
\begin{tabular}{cc|ccccccccccccc}
\hline
\hline
 \diagbox{material}{$\gamma$}& &Bi(x/y) &Bi(z) &  X(x/y) & X(z) &  O(x/y) &O(z)&Cu(x/y)&Cu(z)\\
\hline
Bi$_{2}$O$_{2}$S& & 4.79/4.79 & 4.02 & 4.70/2.40 & 4.93 & 1.30/1.43 & 1.48&-&-\\
Bi$_{2}$O$_{2}$Se& & 4.27/4.27 & 3.42 & 4.05/4.05 & 4.34 & 1.96/1.96 & 1.93&-&-\\
Bi$_{2}$O$_{2}$Te& & 3.60/3.60 & 3.07 & 3.76/3.76 & 5.20 & 1.71/1.71 & 1.50&-&-\\
\hline
BiCuOS && 2.80/2.80&2.37&3.16/3.16 &2.53  &1.99/1.99 &2.0 &2.74/2.74& 3.50   \\
BiCuOSe& & 3.50/3.50 & 3.00  &3.30/3.30 & 2.81 & 1.98/1.98 &1.93& 3.95/3.95 &3.26\\
BiCuOTe& & 3.47/3.47& 2.90 &3.05/3.05& 2.97& 2.26/2.26& 2.06& 3.56/3.56& 2.50\\
\hline
\hline
\end{tabular}
\caption{Projected Gr\"uneisen parameters $\gamma$ on different atoms and directions in BOX and BiCuOX.}
\label{tab:gru}
\end{table*}

\section{Phonon properties of Bi$_2$O$_2$X}
\begin{figure*}[h!]
\includegraphics[scale=0.8]{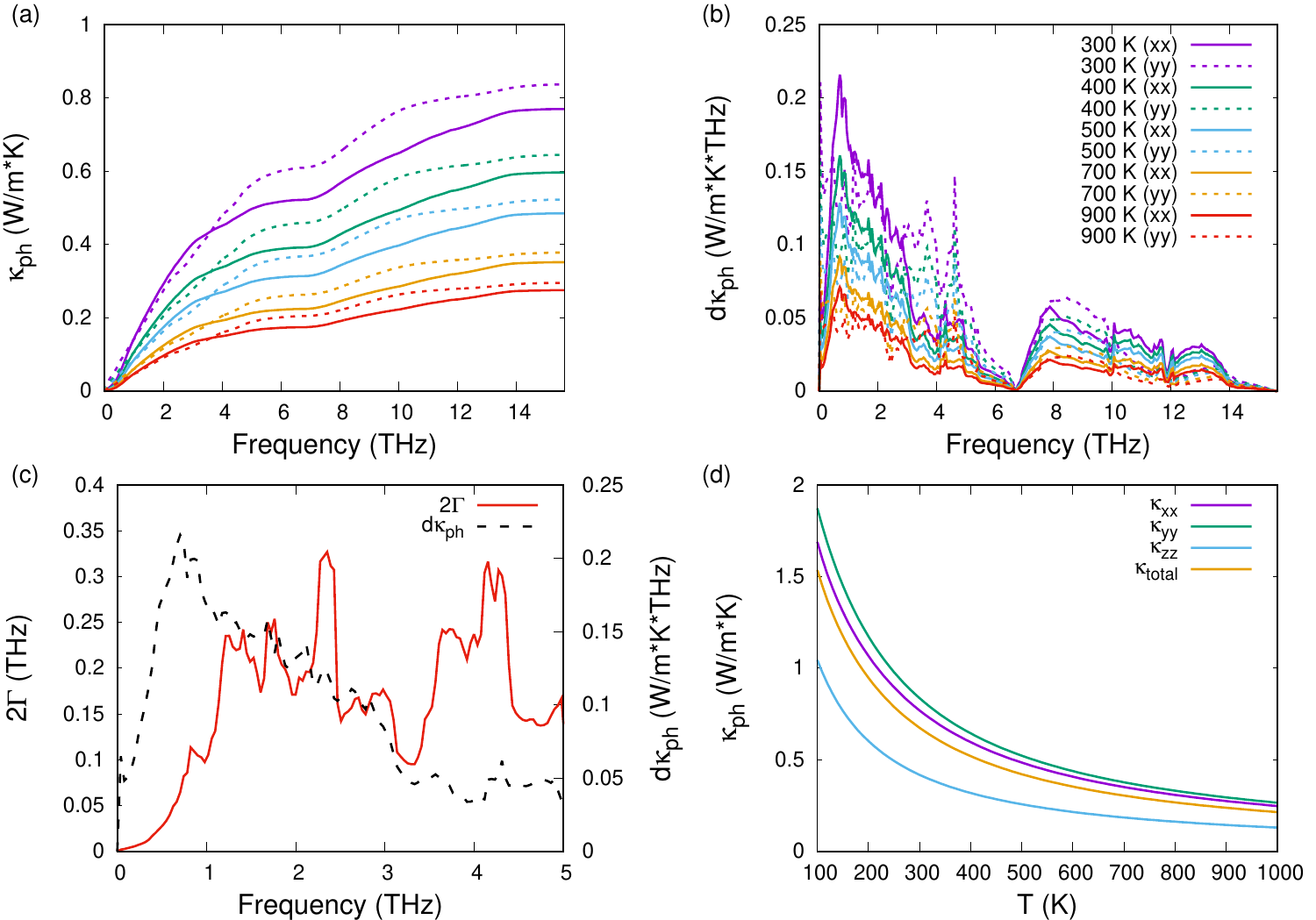}
\caption{The phonon transport coefficients  of bulk  Bi$_{2}$O$_{2}$S. (a) The accumulative phonon conductivity, (b) Frequency dependence of the in-plane modal phonon conductivity at different temperatures. (c) The frequency dependence of modal phonon conductivity  and the phonon line width ($2\Gamma$) at room temperature. (d) The phonon thermal conductivity of bulk Bi$_{2}$O$_{2}$S.}
\label{fig:kBi2O2S}
\end{figure*}

\begin{figure*}[h!]
\includegraphics[scale=0.8]{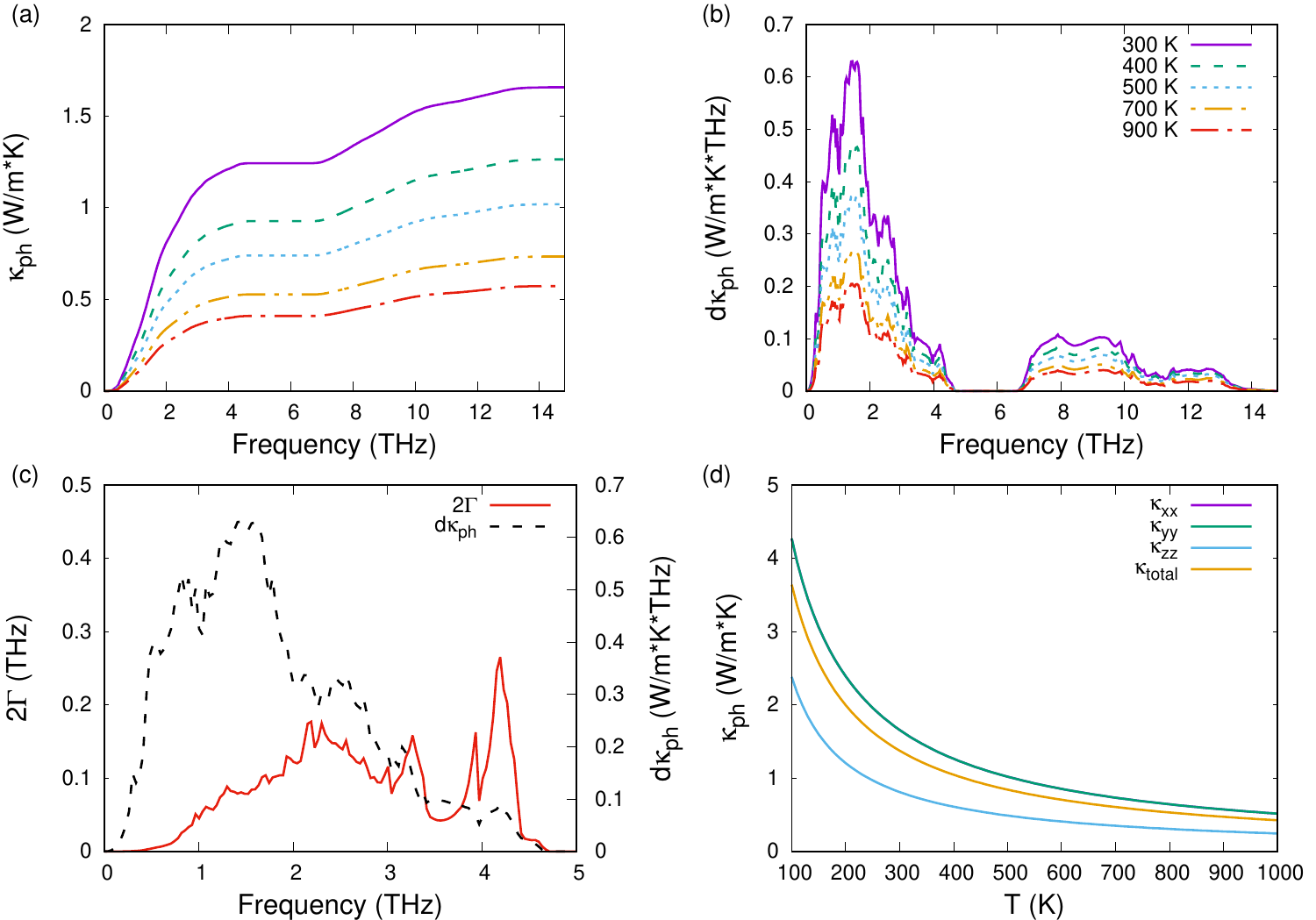}
\caption{Simiar to Fig.~\ref{fig:kBi2O2S} but for Bi$_{2}$O$_{2}$Se.}
\label{fig:kse}
\end{figure*}

\begin{figure*}[h!]
\includegraphics[scale=0.8]{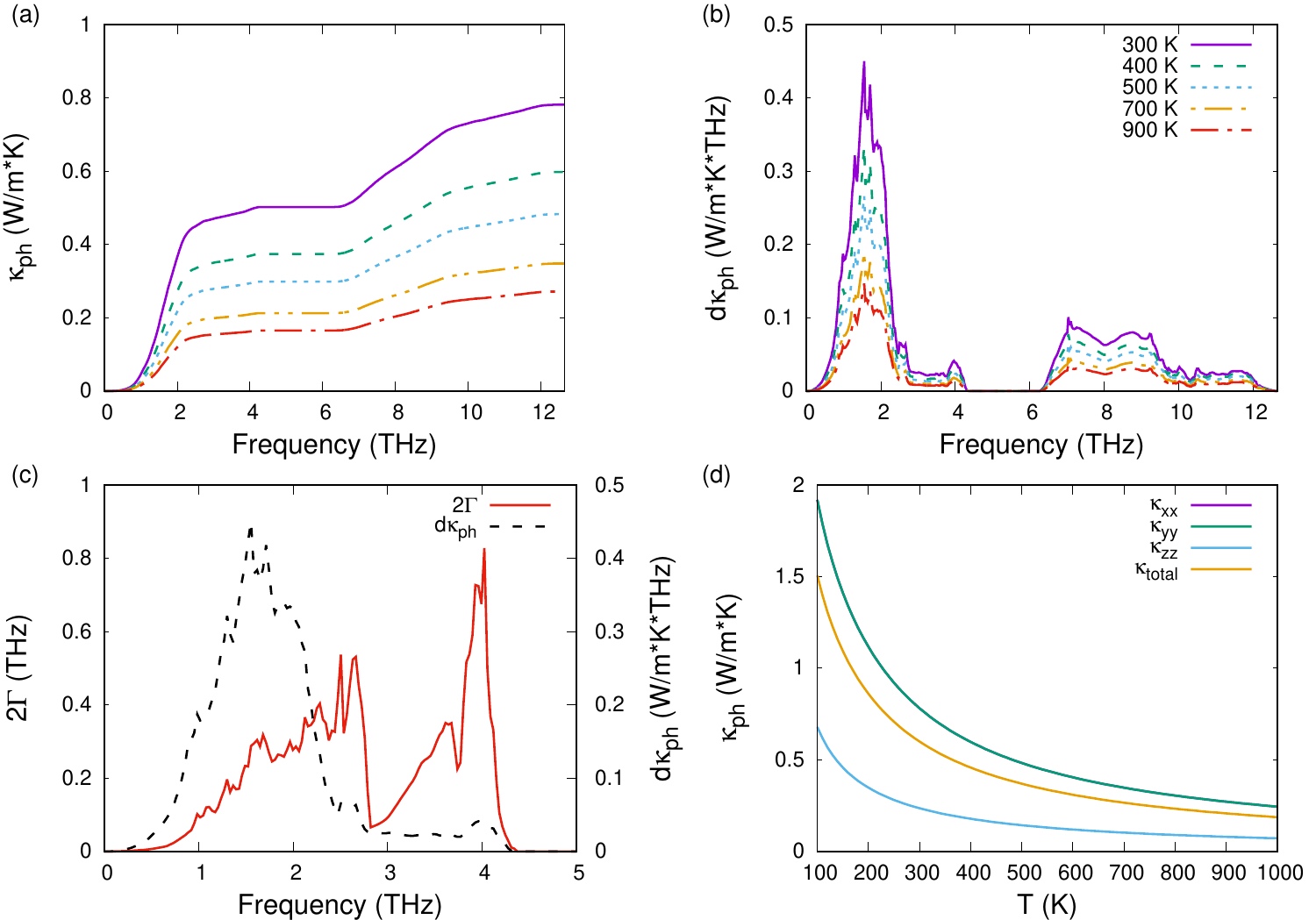}
\caption{Simiar to Fig.~\ref{fig:kBi2O2S} but for Bi$_{2}$O$_{2}$Te.}
\label{fig:kte}
\end{figure*}

Tables~\ref{tab:lattice}-\ref{tab:born} give the calculated lattice parameters, the static dielectric constants and the Born effective charges of Bi$_2$O$_2$X (X=S, Se, Te) (BOX).
Figures~\ref{fig:kBi2O2S}-\ref{fig:kte} give the details of the phonon thermal conductivity ($\kappa$) calculation.

\begin{figure}[!ht]
\includegraphics[scale=0.47]{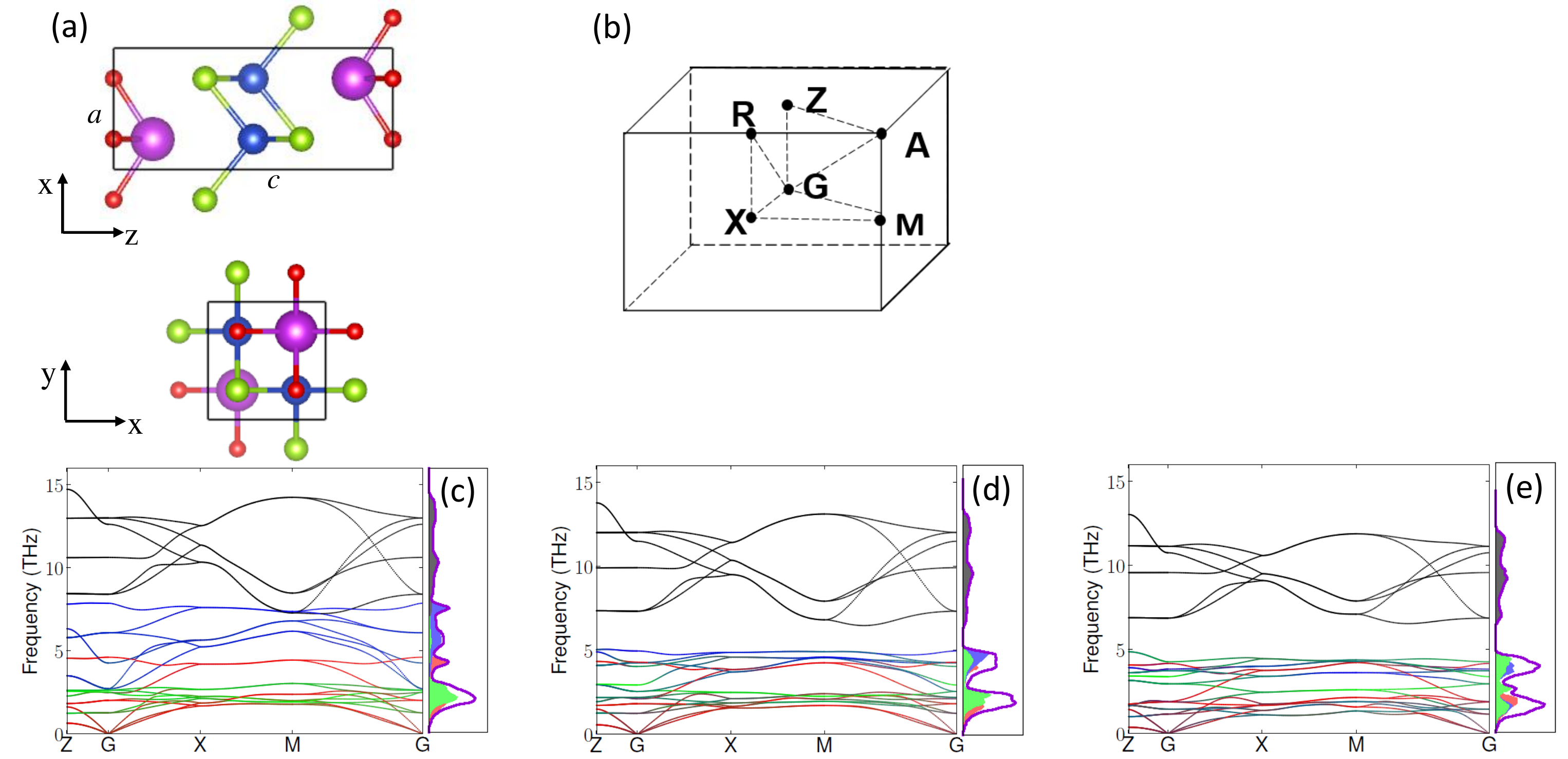}
\caption{(a) Side and top view of the atomic structure of BiCuOX
(X=S, Se, Te). (b) The high symmetry points in the first Brillouin zone:
$\Gamma$ (0.0, 0.0, 0.0), X (0.5 0.0 0.0), M (0.5, 0.5, 0.0 ), A (0.5, 0.5,
0.5), R (0.5, 0.0, 0.5), Z (0.0, 0.0, 0.5). (c-e) The atomically-resolved
phonon dispersion and density of states. The red, blue, black, green colors
correspond to projection onto Bi, chalcogene, O and Cu atoms, respectively.}
\label{fig:cu}
\end{figure}

\begin{figure*}[h!]
\includegraphics[scale=1.0]{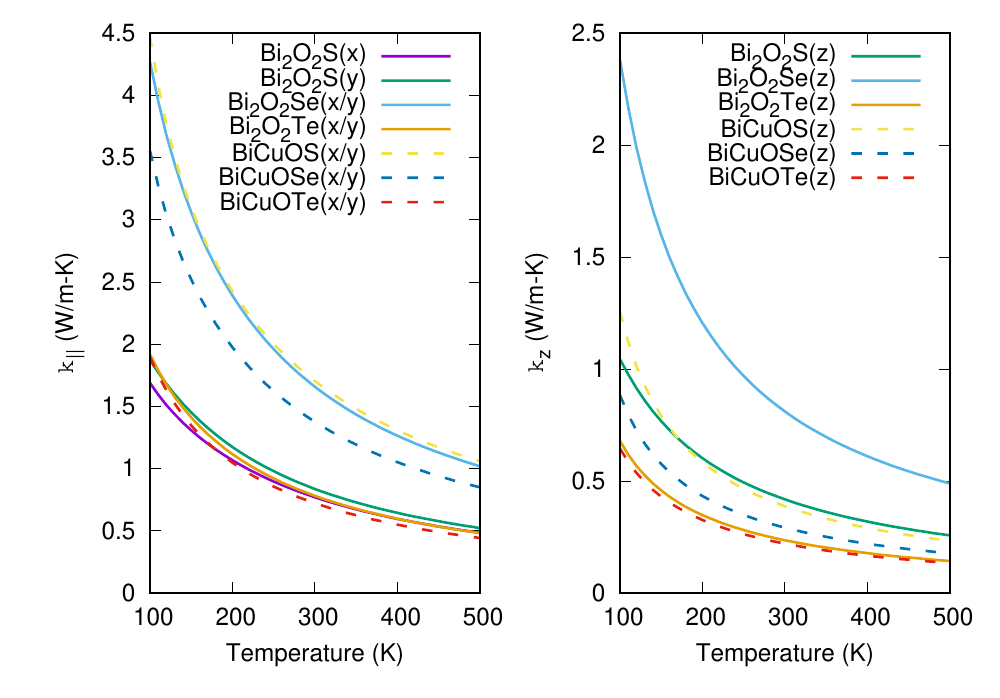}
\caption{Comparison of phonon thermal conductivity as a function of temperature between BOX and BiCuOX (X=S, Se, Te). We find the `abnormal' low value of in-plane thermal conductivity for Bi$_2$O$_2$S . }
\label{fig:kappaall}
\end{figure*}
\begin{figure*}[h!]
\includegraphics[scale=1.5]{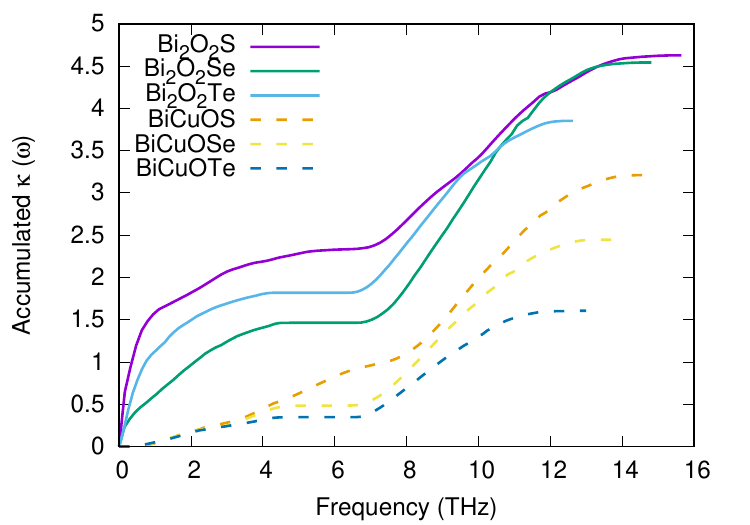}
\caption{ The relative magnitude of phonon band structure contribution to the accumulated phonon thermal conductivity. It is obtained by setting $\tau_\lambda=1$ in Eq.~(1) of the main text.
Comparison between BOX and BiCuOX shows that the former has potential for larger $\kappa$ when only considering the phonon dispersion.
}
\label{fig:knotau2}
\end{figure*}

\section{Comparison to BiCuOX}
BiCuOX (X=S, Se, Te) has similar structure to BOX. Figure~\ref{fig:cu} shows
their lattice structure, corresponding phonon spectrum, and projected density
of states onto different atoms.  $\kappa$ of BiCuOX is also comparable to BOX.
Thus, we here give some comparison between these two kinds of materials.
Figure~\ref{fig:kappaall} shows that, $\kappa$ of BOX and BiCuOX is of similar
magnitude. Figure~\ref{fig:knotau2} shows that, the phonon band structure
contribution to $\kappa$ is different for these two kinds of materials. BOX
would have larger $\kappa$ than BiCuOX if they have the same relaxation time.
Thus, the scattering lifetime $\tau_\lambda$ in BOX is relatively smaller than
that in BiCuOX.

\begin{figure*}[ht!]
\includegraphics[scale=1.5]{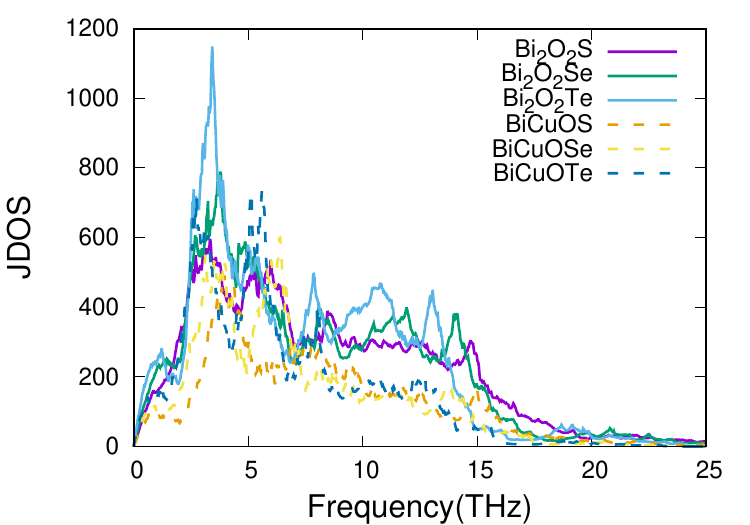}
\caption{The JDOS of BOX and BiCuOX. BOX have larger JDOS than BiCuOX, and hence possiblely larger scattering rate.}
\label{fig:jdo}
\end{figure*}

\begin{figure*}[ht!]
\includegraphics[scale=0.35]{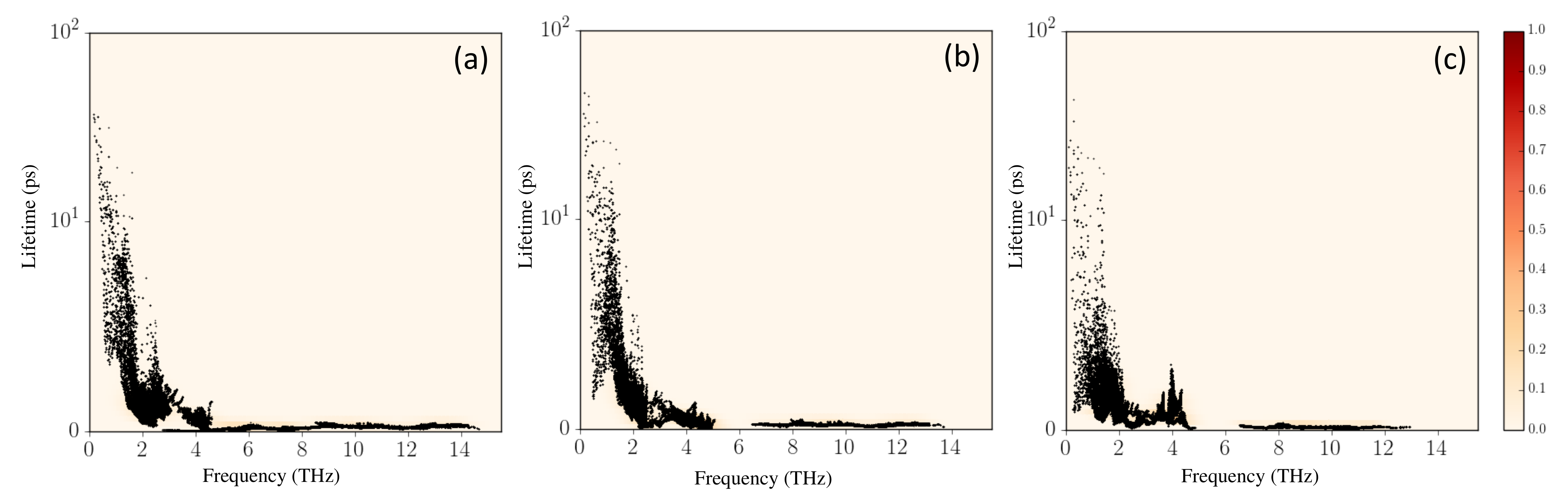}
\caption{The distribution of the scattering lifetime as a function of frequency for BiCuOX.}
\label{fig:life}
\end{figure*}

There are two possible factors that influence $\tau_\lambda$. First,
Figure~\ref{fig:jdo} shows that, the joint density of states (JDOS) for
3-phonon scattering in BOX are larger than that in BiCuOX. Larger JDOS gives
rise to  smaller $\tau_\lambda$. Second, comparison of projected Gr\"uneisen
parameters (Table~\ref{tab:gru}) shows that, the anharmonicity in BOX is larger
than that in BiCuOX.  Especially, the anharmonicity of inter-layer bonding is
stronger in BOX. This can be seen from the relative magnitude of in-plane
($x/y$) and out-of-plane ($z$) $\gamma$ of chalcogen atoms in these two
materials. Summing together, both JDOS and the Gr\"uneisen parameters suggest
that BOX has a shorter scattering lifetime than BiCuOX. This is actually
confirmed in our numerical calculations (Figure 4 in main text and
Figure~\ref{fig:life}).

\clearpage
\bibliography{cite,BOX}

\end{document}